\documentstyle[prl,twocolumn,aps,epsf]{revtex}
\begin{document}

\hoffset-1cm

\draft
\preprint{nucl-th/9610043}

\title{Quantum Mechanical Localization Effects for Bose-Einstein 
Correlations}

\author{U.A. Wiedemann$^{a,b}$, P. Foka$^{b}$, 
H. Kalechofsky$^{b}$, M. Martin$^{b}$, C. Slotta$^{a}$, Q.H. Zhang$^{a}$} 
\address{
   $^a$Institut f\"ur Theoretische Physik, Universit\"at Regensburg,
   D-93040 Regensburg, Germany\\
   $^b$D\'epartment de physique nucl\'eaire et corpusculaire,
   Universit\'e de Gen\`eve, Switzerland\\
}

\date{\today}
\maketitle

\begin{abstract}
For a set of $N$ identical massive boson wavepackets with optimal initial
quantum mechanical localization, we calculate the Hanbury-Brown/Twiss 
(HBT) two-particle correlation function. Our result provides an
algorithm for calculating one-particle spectra and two-particle 
correlations from an arbitrary phase space occupation 
$({\bf q}_i,{\bf p}_i,t_i)_{i=1,N}$ as e.g. returned by 
event generators. It is a microscopic
derivation of the result of the coherent state formalism, providing
explicit finite multiplicity corrections. Both the one- and two-particle 
spectra depend explicitly on the initial wavepacket width $\sigma$
which parametrizes the quantum mechanical wavepacket localization. 
They provide upper and lower bounds
which suggest that a realistic value for $\sigma$ has the order of
the Compton wavelength. 
\end{abstract} 



Two-particle correlations $C({\bf Q},{\bf K})$ of identical
particles are the only known observables giving access to 
the space-time structure of the particle emitting source in
heavy ion collisions. Their interpretation is based on the
result of the coherent state formalism \cite{S73,CH94} which
reads in the plane wave approximation for a large number of
sources
 \begin{mathletters}
   \label{1}
 \begin{eqnarray}
   C({\bf Q},{\bf K}) &=&
   1 + {\left\vert \int d^4x\, S(x,K)\,
   e^{ix{\cdot}Q}\right\vert^2 \over
   \int d^4x\, S(x,P_1)\, \, \int d^4y\, S(y,P_2) } \, ,
   \label{1a} \\
   Q &=& P_1 - P_2\, ,
   \qquad K = \textstyle{1\over 2}(P_1 + P_2)\, .
 \label{1b}
 \end{eqnarray}
 \end{mathletters}
In this setting, an Hanbury-Brown/Twiss (HBT) interferometric analysis
aims at extracting from the correlator $C({\bf Q},{\bf K})$ as much 
information as possible about the space-time emission function
$S(x,K)$. Since this emission function cannot
be reconstructed unambiguously from $C({\bf Q},{\bf K})$ \cite{H96}, 
(\ref{1}) is mainly used in the study of model emission functions 
$S(x,K)$. These studies have
clarified to a considerable extent the question which geometrical
and dynamical source characteristics are reflected in which
particular momentum dependencies of the correlator (cf. \cite{H96}
and refs. therein). A comparison
with measured correlations then allows to constrain the class
of source models consistent with data. 

Microscopic event generators are one important tool to generate model 
emission functions. Here, we do not discuss in
how far existing event generators (e.g. RQMD~\cite{SSG89}, 
VENUS~\cite{W93}, ARC~\cite{PSK92}) provide an internally consistent
calculation of the phase space distribution. None of them 
propagates (anti)-symmetrized N-particle states from first 
principles, and the resulting difficulties in calculating
$2$-particle correlations have been discussed recently in great 
detail \cite{A96}. The typical event generator output is a set 
$\Sigma$ of phase space points at given times 
$z_i = ({\bf q}_i,{\bf p}_i,t_i)$ 
which one associates with the ``points of last interactions''.
However, the Heisenberg uncertainty principle allows to interpret
the $z_i$ only as mean positions of boson wave packets. To specify
the localization of these wavepackets in phase space, at least
one additional parameter is needed, e.g. the initial spatial wavepacket
width $\sigma$. 
Irrespective of how the phase space occupation has been
obtained, we shall take the set $\Sigma$ and the width $\sigma$
as initial condition for the present investigation: 
$\Sigma$ and $\sigma$ {\it define}
the boson emitting source. For notational simplicity, we restrict 
our discussion to one particle species, negative pions, say. 

The problem in associating an emission function 
$S(x,K)$ to the distribution $\Sigma$ is that $\Sigma$ 
is a {\it discrete} phase space distribution of {\it on-shell} 
particles. In contrast, the emission function $S(x,K)$ of the 
coherent state formalism is a {\it continuous} distribution which 
allows for {\it off-shell} momenta $K$. Often, one circumvents this 
problem by approximating the Pratt algorithm~\cite{P90} by an ad hoc
prescription: each particle pair $(i,j)$ is weighted 
with a probability $\rho_{ij}$, $q_i$ being 4-vectors 
$q_i=(t_i,{\bf q}_i)$,
 \begin{mathletters}
  \label{2}
  \begin{eqnarray}
  \label{2a}
    C({\Delta\bf Q},{\Delta\bf K}) &=&    
    \textstyle{1\over N({\Delta\bf Q},{\Delta\bf K})}
    \sum_{(i,j)}\, \rho_{ij}\, , \\
    \label{2b}
    \rho_{ij} &=& 1 + \cos( (p_i-p_j)\cdot (q_i-q_j))\, .
  \end{eqnarray}
  \end{mathletters}
Here, $C({\Delta\bf Q},{\Delta\bf K})$ denotes the 2-particle 
correlator for pairs whose relative and average pair 
momenta ${\bf p}_i- {\bf p}_j$, $\textstyle{1\over 2}({\bf p}_i+ {\bf p}_j)$
lie in the bin $\Delta \bf Q$, $\Delta \bf K$.
$N({\Delta\bf Q},{\Delta\bf K})$ is the corresponding number of 
particle pairs. A tentative argument to justify the prescription
(\ref{2}) is that $\rho_{ij}$ coincides with the formal Born 
probability density ${\Psi}^*\Psi$ of the Bose-Einstein 
symmetrized 2-particle plane wave
 \begin{mathletters}
  \label{3}
  \begin{eqnarray}
    \label{3a}
    \rho_{ij} &=& \Psi^*(q_i,q_j,p_i,p_j)\, \Psi(q_i,q_j,p_i,p_j)\, ,\\
    \label{3b}
    \Psi(q_i,q_j,p_i,p_j) &=& {1\over \sqrt{2}}
    {\left({ {\rm e}^{ip_iq_i + ip_jq_j} 
          + {\rm e}^{ip_jq_i + ip_iq_j}
          }\right)}\, .
  \end{eqnarray}
 \end{mathletters}
However, the prescription (\ref{2}) based on the ansatz (\ref{3}) is 
inconsistent \cite{MKF96} with the result (\ref{1}) of the coherent state
formalism: The correlator in (\ref{1}) is always larger than 
unity \cite{H96}. In contrast, the expression
(\ref{2}) can drop below unity in the region of sufficiently large 
relative momenta \cite{MKF96}. 
Also, the prescription (\ref{2}) is difficult to reconcile with 
quantum mechanical localization requirements since the
plane wave (\ref{3b}) cannot be an eigenstate for both the position
and momentum operator.

In what follows, we take the quantum mechanical localization of
bosons into account by associating to the phase space emission
points $z_i$ free Gaussian wavepackets of initial spatial width 
$\sigma$, \cite{PGG90,MP97}
  \begin{mathletters}
  \label{4}
  \begin{eqnarray}
   f_{z_i}^{(\sigma)}({\bf X},t) 
    &=& (\pi \sigma^2)^{-\textstyle{3\over 4}}
    {\left({ \sigma^2 \over \sigma^2_i(t) 
          }\right)}^{\textstyle{3\over 2}}
    \exp{\left({ i{\bf p}_i{\bf X} - i{\cal E}_it}\right)}
    \nonumber \\
    && \times \exp{\left({-\textstyle{1\over 2\sigma^2_i(t)}\, 
      ({\bf X}-{\bf q}_i(t))^2 }\right)}\, 
  \label{4a} \\
   {\bf q}_i(t) &=& {\bf q}_i 
                    + \textstyle{{\bf p}_i\over m}(t-t_i)\, ,
   \qquad {\cal E}_i = {{\bf p}_i^2\over 2m}\, ,
  \label{4b} \\
  \sigma^2_i(t) &=& \sigma^2 + i\,\textstyle{{(t-t_i)}\over m}\, .
  \label{4c}
  \end{eqnarray}
  \end{mathletters}
This one boson state (\ref{4a}) is optimally
localized around $({\bf q}_i,{\bf p}_i)$ in the sense that it saturates the
Heisenberg uncertainty relation 
$\Delta {\bf x}\cdot \Delta {\bf p}_x = 1$,
with $\Delta x_i = \sigma$ at time $t=t_i$.
The time evolution of (\ref{4}) is the free
unperturbed evolution determined by the Hamiltonian $H_0 = {\Delta\over 2m}$,
$\Delta$ being the Laplacian. Since the $i$-th and 
$j$-th boson are identical, we associate to the
two emission points $z_i$ and $z_j$ the 
symmetrized two boson wave function $\Phi_{ij}({\bf X},{\bf Y},t)$
(the normalization factor is omitted and plays no role in what follows)
  \begin{eqnarray}
    \Phi_{ij}({\bf X},{\bf Y},t)
    &=& f_{z_i}^{(\sigma)}({\bf X},t) f_{z_j}^{(\sigma)}({\bf Y},t)
    \nonumber \\
    && + f_{z_i}^{(\sigma)}({\bf Y},t) 
                f_{z_j}^{(\sigma)}({\bf X},t) \, .
    \label{5}
  \end{eqnarray}
We now derive an algorithm for calculating one-particle spectra
$\nu({\bf P})$ and two-particle correlations $C({\bf P}_1,{\bf P}_2)$
from an arbitrary initial phase space distribution $\Sigma$ of
best localized boson wavepackets $f_{z_i}^{(\sigma)}$.
Our first step is to calculate for two identical bosons 
the detection probability at time $t$ at the positions 
${\bf X}$ and ${\bf Y}$ with momenta ${\bf P}_1$, ${\bf P}_2$ respectively. 
This is given by the two-particle Wigner phase space density~\cite{HOSW84} 
  \begin{eqnarray}
   &&W_{ij}({\bf X},{\bf Y},{\bf P}_1,{\bf P}_2,t)
     = \Phi_{ij}({\bf X},{\bf Y},t) (2\pi)^6 
                       \delta^{(3)}({\bf P}_1 - \hat{\bf P}_1)
   \nonumber \\
   && \qquad\qquad \qquad \qquad \times 
         \delta^{(3)}({\bf P}_2 - \hat{\bf P}_2)
         \Phi_{ij}^*({\bf X},{\bf Y},t)
                       \nonumber \\
   &&\, = \int d^3{\bf X}_1\, d^3{\bf Y}_1 
         \Phi({\bf X} + \textstyle{ {\bf X}_1\over 2},
         {\bf Y} + \textstyle{ {\bf Y}_1\over 2},t)\,
         {\rm e}^{i{\bf P}_1{\bf X}_1}\, 
   \nonumber \\
   && \qquad \qquad \qquad \times {\rm e}^{i{\bf P}_2{\bf Y}_1}\, 
         \Phi^*({\bf X} - \textstyle{ {\bf X}_1\over 2},
         {\bf Y} - \textstyle{ {\bf Y}_1\over 2},t)\, .
    \label{6}
  \end{eqnarray}
The corresponding probability to detect these bosons with momenta
${\bf P}_1$ and ${\bf P}_2$ irrespective of their position is 
  \begin{mathletters}
    \label{7}
  \begin{eqnarray}
    P_{ij}({\bf P}_1,{\bf P}_2) &=&
        \int d^3{\bf X}\, d^3{\bf Y}\, 
        W_{ij}({\bf X},{\bf Y},{\bf P}_1,{\bf P}_2,t) 
     \nonumber \\
       &=&  w_i({\bf P}_1,{\bf P}_1)\, w_j({\bf P}_2,{\bf P}_2)
     \nonumber \\
        &&+ w_i({\bf P}_2,{\bf P}_2)\, w_j({\bf P}_1,{\bf P}_1)
     \nonumber \\
    &&+ 2\, w_i({\bf P}_1,{\bf P}_2)\, w_j({\bf P}_1,{\bf P}_2)
     \nonumber \\
     && \qquad \times \cos {\left({ (q_i-q_j)\cdot (P_1-P_2)}\right)}\, ,
     \label{7a} \\
      w_i({\bf P}_1, {\bf P}_2) &=&
        {\rm e}^{-\textstyle{\sigma^2\over 4} 
                 ({\bf P}_1- {\bf P}_2)^2}\, 
               s_i({\bf K})\, ,
      \label{7b} \\
      s_i({\bf K}) &=&  2^3\, (\pi\sigma^2)^{3\over 2}\, 
         {\rm e}^{- \sigma^2 
         {\left({ {\bf p}_i - {\bf K}}\right)}^2}\, .
       \label{7c}
   \end{eqnarray}
   \end{mathletters}
Here, $P_i$ denotes 4-vectors 
$P_i = (\textstyle{1\over 2m}{\bf P}_i^2, {\bf P}_i)$.
We note that $P_{ij}$ is independent 
of the detection time $t$, i.e., only the correlations which exist
already at emission are measured at time $t$ in the detector.
This $t$-independence is a consequence of the free time evolution; 
it is lost if final state interactions are included
in the evolution of the wavepackets (\ref{4}).
Neglecting higher order symmetrizations, we define the (unnormalized)
two pion correlation $R({\bf P}_1,{\bf P}_2)$ for a set of $N$ phase 
space points $z_i$ by summing the probabilities $P_{ij}$ over
all $\textstyle{1\over 2}N(N-1)$ pairs $(i,j)$
  \begin{mathletters}
        \label{8}
  \begin{eqnarray}
   &&R({\bf P}_1,{\bf P}_2) = \sum_{(i,j)} P_{ij}({\bf P}_1,{\bf P}_2)\, .
    \nonumber \\
   && \qquad = \nu({\bf P}_1)\, \nu({\bf P}_2) - 2\, T_c({\bf P}_1,{\bf P}_2)
    \nonumber \\
   && \qquad + \Big\vert \sum_{i=1}^N w_i({\bf P}_1,{\bf P}_2)
        {\rm e}^{it_i(E_1-E_2) - 
          i{\bf q}_i({\bf P}_1-{\bf P}_2)}\Big\vert^2\, ,
   \label{8a}\\
   && \nu({\bf P}) =
        \sum_{i=1}^N s_i({\bf P})\, .
        \label{8b}
  \end{eqnarray}
  \end{mathletters}
Here, $s_i({\bf P})$ is the one-particle probability (\ref{7c}) that
a boson in the state $f_{z_i}^{(\sigma)}$ is detected 
with momentum ${\bf P}$. Accordingly, $\nu({\bf P})$ 
is the one-particle spectrum of the distribution
$\Sigma$ with spatial localization $\sigma$. 
The contribution $T_c$ to $R({\bf P}_1,{\bf P}_2)$ corrects for the 
fact that the sums in the other two terms of (\ref{8a}) include
the $N$ identical pairs $(i,i)$ which are not present in 
$R({\bf P}_1,{\bf P}_2)$, 
  \begin{equation}
    T_c({\bf P}_1,{\bf P}_2) = 
         \sum_{i=1}^N s_i({\bf P}_1)\, s_i({\bf P}_2)\, .
  \label{9}
  \end{equation}
To obtain a normalized 2-particle correlation $C({\bf P}_1,{\bf P}_2)$, 
we choose the normalization 
  \begin{mathletters}
    \label{10}
  \begin{eqnarray}
    N({\bf P}_1,{\bf P}_2) &=& \nu({\bf P}_1)\, \nu({\bf P}_2)
    - T_c({\bf P}_1,{\bf P}_2)\, ,
    \label{10a} \\
    C({\bf P}_1,{\bf P}_2) &=& { R({\bf P}_1,{\bf P}_2) \over 
                               N({\bf P}_1,{\bf P}_2)}\, .
  \label{10b}
  \end{eqnarray}
  \end{mathletters}
This choice is motivated by the experimental
praxis of ``normalization by mixed pairs'': An uncorrelated 
(mixed) pair is described by an unsymmetrized product state
  \begin{equation}
    \Phi_{ij}^{\rm uncorr}({\bf X},{\bf Y},t)
    = f_{z_i}^{(\sigma)}({\bf X},t) f_{z_j}^{(\sigma)}({\bf Y},t)\, ,
    \label{11}
  \end{equation}
for which the two particle Wigner phase
space density and the corresponding detection probability
$P_{ij}^{\rm uncorr}({\bf P}_1,{\bf P}_2)$ can be calculated 
according to (\ref{6}). Taking both
distinguishable states $\Phi_{ij}^{\rm uncorr}$ and $\Phi_{ji}^{\rm uncorr}$
into account, and summing over all pairs $(i,j)$, we find
  \begin{equation}
    N({\bf P}_1,{\bf P}_2) = \sum_{(i,j)} 
    P_{ij}^{\rm uncorr}({\bf P}_1,{\bf P}_2)\, .
  \label{12}
  \end{equation}
Hence, the normalization (\ref{10a}) is
the 2-particle detection probability for uncorrelated pairs.
The correlator reads 
  \begin{equation}
    C({\bf P}_1,{\bf P}_2) = 1 + 
    {{ \Big\vert \sum_{i=1}^N w_i({\bf P}_1,{\bf P}_2)
        {\rm e}^{iq_iQ}\Big\vert^2
        - T_c}
      \over
         {\nu({\bf P}_1)\, \nu({\bf P}_2) - T_c } }\, .
   \label{13}
  \end{equation}
In contrast to (\ref{2}), this is a continuous function of the measured
momenta $P_1$, $P_2$, i.e., no binning of the correlator is 
required. Note that the normalization (\ref{10a}) ensures that
the correlator (\ref{13}) is always smaller than $2$ and equals
$2$ for ${\bf P}_1-{\bf P}_2=0$. (This follows from the fact 
that the sum of the first two 
terms in (\ref{7a}) is always larger than the third one.)
For any boson source, defined by an arbitrary phase space
distribution $\Sigma$ and a spatial wavepacket width 
$\sigma$, Eq. (\ref{13}) provides an algorithm of how
to calculate the 2-particle correlator, using
(\ref{7b}), (\ref{7c}), (\ref{8b}) and (\ref{9}).

To understand how the correlator (\ref{13}) 
relates to the result of the coherent state formalism (\ref{1}),
the limit of a large number $N$ of emission points is relevant.
$T_c({\bf P}_1,{\bf P}_2)$ 
in (\ref{13}) is a sum over $N$ terms while the other terms in 
the nominator and denominator are sums of $N^2$ terms. 
In this sense, the $T_c$-dependence of (\ref{13}) provides a
finite multiplicity correction and  
can be neglected as a subleading $\textstyle{1\over N}$-contribution
in the large $N$ limit of (\ref{13}), 
  \begin{equation}
    \lim_{N\to \infty} C({\bf P}_1,{\bf P}_2) = 1 + 
    {{ {\rm e}^{-\textstyle{\sigma^2\over 2}{\bf Q}^2}\,
        \Big\vert \sum_{i=1}^\infty s_i({\bf K})
        {\rm e}^{iq_iQ}\Big\vert^2}
      \over
      { {\left({ \sum_{i=1}^\infty s_i({\bf P}_1)}\right)}\, 
      {\left({ \sum_{j=1}^\infty s_j({\bf P}_2)}\right)} } }\, . 
   \label{14}
  \end{equation}
We note that in the derivation of (\ref{1}), subleading 
$\textstyle{1\over N}$-contributions are dropped~\cite{CH94}. 
The large $N$ approximations (\ref{1}) and (\ref{14}) 
are clearly justified for pion interferometry in 
ultrarelativistic (Pb-Pb) heavy ion collisions
where typical pion multiplicities are in the hundreds.
For smaller systems, however, and especially in
studies of the multiplicity dependence of HBT correlations
\cite{OPAL}, one might wish to start from the expression
for finite multiplicity (\ref{13}). 
 Expression (\ref{14}) can be obtained from
the coherent state result (\ref{1}) by inserting 
  \begin{mathletters}
    \label{15}
  \begin{eqnarray}
    S(x,K) &=& \sum_{i=1}^\infty S_i(x,{\bf K})\, ,
    \label{15a}\\
    S_i(x,{\bf K}) &=& {\cal N}\, \delta(t-t_i)\,
    {\rm e}^{-\textstyle{1\over \sigma^2} ({\bf x}-{\bf q}_i)^2}
    \nonumber \\
    && \times {\rm e}^{-\textstyle{\sigma^2} ({\bf K}-{\bf p}_i)^2}\, ,
    \label{15b}
  \end{eqnarray}
  \end{mathletters}
where ${\cal N}$ is an arbitrary normalization factor.
In this sense, (\ref{15a}) is the emission function for a source
$\Sigma$ with initial spatial localization $\sigma$. It contains 
the information about how the {\it initial} phase space emission 
points $z_i$ and the {\it measured} momenta ${\bf K}$ are correlated.
Spatial and temporal components are not treated equally in
(\ref{15b}), since our derivation is not Lorentz covariant. The
Lorentz covariant setting used in (\ref{1}) 
allows for an additional dependence of $S(x,K)$ on the 
temporal component of $K$ which does not exist in our  
derivation. In practical applications however, the emission function 
(\ref{1}) is used in the so-called on-shell approximation, where this 
additional $K_0$-dependence is not employed, \cite{H96}.

Both the 2-particle correlator (\ref{13}) and the 1-particle spectrum 
$\nu({\bf P})$ in (\ref{8b}) depend on the initial spatial 
localization $\sigma$ which is an additional free parameter.
We now discuss this $\sigma$-dependence.
We first consider the limit $\sigma \to 0$, in which the Gaussian 
wavepacket (\ref{4a}) describes at freeze out ($t = t^{(i)}$) a state
with position uncertainty $\Delta {\bf x} = 0$, i.e., the source is sharply
(``classically'') localized in configuration space. The prize for this
optimal spatial information is that nothing can be said about the 
initial momenta ${\bf p}_i$ at emission, the one-particle spectrum 
$\nu({\bf P})$ is flat. The 
measured momentum correlations contain spatial information 
about the source, namely
  \begin{equation}
    \lim_{\sigma\to 0} C({\bf P}_1,{\bf P}_2) = 1 +
    { \sum_{(i,j)} \cos{\left({ (q_i-q_j)\cdot (P_1-P_2)}\right)}
      \over N(N-1)}\, .
    \label{16}
  \end{equation}
Due to the $\cos$-term, the dependence of
the 2-particle correlator (\ref{16}) on the {\it measured} relative momentum
$P_1 - P_2$ gives information on the {\it initial} relative distances 
$q_i - q_j$ in the source. This is the HBT effect.
Eq. (\ref{16}) differs significantly from the 
$\cos$-prescription (\ref{2}): here, $P_1 - P_2$ is the measured 
relative pair momentum, while $p_i - p_j$ in (\ref{2}) denotes the 
initial momentum difference. As a consequence, 
the sum $\sum_{(i,j)}$ in (\ref{16}) goes over {\it all} 
pairs irrespective of the momenta $p_i$, $p_j$ since in the limit 
$\sigma \to 0$, all information about these initial momenta is lost, 
while the sum in (\ref{2}) goes only over those pairs for which 
the initial relative pair momentum $p_i - p_j$ lies in the same bin as 
the measured $P_1 - P_2$. Since the correlator (\ref{16}) is a limiting
case of (\ref{14}), it is always larger than unity. In contrast, due
to the wrong pair selection criterion, the correlator (\ref{2}) can
drop below $1$. This insufficiency of (\ref{2}) becomes more
significant for sources with strong $q$-$p$ position-momentum correlation,
as was noticed in \cite{MKF96}.

The other limiting case of (\ref{13}) is the plane wave limit
  \begin{equation}
    \lim_{\sigma\to \infty} C({\bf P}_1,{\bf P}_2) = 1 +
           \delta_{{\bf P}_1,{\bf P}_2}\, .
    \label{17}
  \end{equation}
In this limit, nothing can be said about the spatio-temporal extension 
of the source since the 2-particle symmetrized wave functions (\ref{5}) 
contain no space-time information. 

The difference between (\ref{16}) and (\ref{17}) 
shows that the $\sigma$-dependence
of the $2$-particle correlator cannot be neglected. 
As pointed out already in \cite{PGG90,MP97}, none of the two 
limits is realistic. For $\sigma \to 0$, one has sharp information 
in configuration space but the momentum space information is lost and 
hence, the set $\Sigma$ of phase space emission points 
$({\bf q}_i,{\bf p}_i,t_i)$ contains no information about the 
one-particle momentum spectrum $\nu({\bf P})$. In the limit 
$\sigma \to \infty$, on the other hand, no space-time information 
is contained in $\Sigma$. A realistic width $\sigma$ hence lies in between 
these two extremes. 

For further discussing the $\sigma$-dependence of 1- and 2-particle
spectra, we now generalize Eq.\ (\ref{15}) to continuous phase space 
distributions  $\rho({\bf q},{\bf p}, t)$ which can encode statistical
assumptions. To this aim, we extend the sum (\ref{15a}) 
to a phase space integral weighted by a (classical) distribution 
$\rho$,~\cite{CH94,SWZH}
  \begin{equation}
    S(x,K) = \int d^3{\bf q}_i\, d^3{\bf p}_i\, dt_i\,
    \rho({\bf q}_i,{\bf p}_i,t_i)\, S_i(x,{\bf K})\, .
    \label{18}
  \end{equation} 
The one-particle spectrum is obtained from (\ref{18}) via 
$\int d^4x\, S(x,K)$. Especially, for a Boltzmann distribution of
temperature $T$, $\rho({\bf q},{\bf p},t) \propto \exp(
-\textstyle{ {\bf p}^2\over 2mT})$, this
one-particle spectrum has an effective temperature~\cite{MP97}
  \begin{equation}
    T_{\rm eff} = T + \textstyle{1\over 2m\sigma^2}\, .
  \label{19}
  \end{equation}
Hence, the one-particle spectrum broadens significantly for
a narrow spatial width $\sigma$.
$\sigma$ is a free parameter which has to be determined
from a comparison to data. How can this be done? One idea is to look 
at systems which can be expected to provide very small, almost pointlike 
boson emission regions. Candidates are e.g. the $p$-$\bar{p}$ annihilation 
process \cite{GGLP} or $Z_0$-decays \cite{OPAL}. The width of the 
HBT-correlator determined for these systems should be dominated by 
the width $\sigma$. To obtain an argument supporting this idea,
we consider the extreme case of a ``pointlike source'' $\rho$ with no
momentum dependence, for which all particles are emitted from the
same space-time position $\tilde{\bf q}$, $\tilde{t}$. Calculating
the emission function (\ref{18}) for the corresponding  
$\rho({\bf q}_i,{\bf p}_i,t_i) = \delta^{(3)}({\bf q}_i - \tilde{\bf q})
\delta(\tilde{t}-t_i)$, we find
  \begin{mathletters}
    \label{20}
  \begin{eqnarray}
    C({\bf P}_1,{\bf P}_2) &=& 1 + 
    {\rm e}^{-\textstyle{\sigma^2\over 2} {\bf Q}^2}\, ,
   \label{20a} \\
    R_{\rm point}^{\rm HBT} &=& \sigma/\sqrt{2} \, .
  \label{20b}
  \end{eqnarray}
  \end{mathletters}
Several assumptions enter this result: for pointlike sources with an
additional momentum dependence, the correlation is in general
more complicated. Also, the width $\sigma$ could in principle depend
on the emission points $z_i$, the localized wavepackets could have 
a different, non-Gaussian shape, etc. Still, Eq.~(\ref{20}) suggests that the
size of the HBT radius parameters measured for very small boson emitting
systems is essentially given by $\sigma$. 

From the pion interferometric 
measurements of systems like the $p$-$\bar{p}$ annihilation process or 
$Z_0$-decays \cite{GGLP,OPAL}, one infers on the
basis of (\ref{20b}) a pion wavepacket width of the order 
$\sigma \approx 1$ fm. This is in good agreement with the natural
localization scale of the pion, its Compton wavelength~\cite{GGLP}. 
Remarkably, for such a localization, the additional quantum contribution to 
the temperature $T_{\rm eff}$ in (\ref{20})  is of order
$\textstyle{1\over 2m\sigma^2} \approx 100$ MeV. This indicates that
the initial spatial localization width $\sigma$ plays an important 
role in accounting for the slope of the measured one-particle spectra.

In the present formalism, the role of an event generator for the boson
emitting source is to provide a dynamical calculation of the phase
space occupation $\Sigma$ from some more fundamental initial condition.
The current praxis for event generators of heavy ion collisions amounts
to determining the $1$-particle spectrum in the limit $\sigma \to \infty$.
We have shown that this limit is unrealistic and that a realistic spatial
wavepacket width leads to a substantial broadening of the one-particle
spectrum. Our main result is an algorithm which allows for the
calculation of both the one-particle spectra via $\nu({\bf P})$ in 
(\ref{8b}), and the two-particle correlations via 
$C({\bf P}_1,{\bf P}_2)$ in (\ref{13}), starting from
an arbitrary initial phase space distribution $\Sigma$ 
of wavepackets with arbitrary spatial localization $\sigma$.
Here, the spatial width $\sigma$ is an additional free parameter
which has to be fixed in comparison with experimental data. The
slope of the transverse mass spectra (\ref{19}) provide a lower bound
for $\sigma$ while the 2-particle correlators provide an upper bound.
For pions these bounds are very tight and a 
realistic width $\sigma$ is of the order of the pion
Compton wavelength.

We thank U. Heinz for helpful remarks and a critical reading of the
manuscript. Stimulating discussions with J.-P. Naef and L. Rosselet
are acknowledged. This work was supported by BMBF, DFG, DPNC and GSI. 

\end{document}